\documentclass[conference]{IEEEtran}
\IEEEoverridecommandlockouts

\usepackage{multirow}%
\usepackage{amsthm}%
\usepackage{amsmath}
\usepackage{mathrsfs}%
\usepackage{tabularx} 

\usepackage{hyperref}
\usepackage{xurl}

\usepackage[table,xcdraw]{xcolor}

\usepackage{url}

\usepackage{textcomp}%
\usepackage{lmodern}

\usepackage{cite}
\usepackage{amsmath,amssymb,amsfonts}
\usepackage{algorithmic}
\usepackage{graphicx}
\usepackage{textcomp}

\def\BibTeX{{\rm B\kern-.05em{\sc i\kern-.025em b}\kern-.08em
    T\kern-.1667em\lower.7ex\hbox{E}\kern-.125emX}}
    
\begin{document}


\title{HAL 9000: a Risk Manager for ITSs
\\
\thanks{
This work is financed by National Funds through the Portuguese funding agency, FCT - Fundação para a Ciência e a Tecnologia, within project UIDB/50014/2020.
DOI 10.54499/UIDB/50014/2020 https://doi.org/10.54499/uidb/50014/2020
This work was funded by 2021.08532.BD (FCT), and by 2021.04529.BD (FCT).}
}


\author{
Tadeu Freitas\IEEEauthorrefmark{1}\IEEEauthorrefmark{2}, Carlos Novo\IEEEauthorrefmark{1}, João Soares\IEEEauthorrefmark{1}\IEEEauthorrefmark{2}, Inês Dutra\IEEEauthorrefmark{1}\IEEEauthorrefmark{3},\\ Manuel E. Correia\IEEEauthorrefmark{1}\IEEEauthorrefmark{2}, Behnam Shariati\IEEEauthorrefmark{4}, Rolando Martins\IEEEauthorrefmark{1}\IEEEauthorrefmark{5}\\
\IEEEauthorblockA{\IEEEauthorrefmark{1}Faculty of Sciences, University of Porto, Portugal}
\IEEEauthorblockA{\IEEEauthorrefmark{2}CRACS/INESC-TEC, Portugal}
\IEEEauthorblockA{\IEEEauthorrefmark{3}CINTESIS@RISE, Portugal}
\IEEEauthorblockA{\IEEEauthorrefmark{4}University of Maryland, Baltimore County, USA}
\IEEEauthorblockA{\IEEEauthorrefmark{5}SafeHelm, lda, Porto, Portugal\\
\{tadeufreitas, joao.soares, mdcorrei, carlosnovo, ines\}@fc.up.pt,\\ shariati@umbc.edu, rmartins@safehelm.com}
}

\maketitle

\begin{abstract}

HAL 9000 is an Intrusion Tolerant Systems (ITSs) Risk Manager, which assesses configuration risks against potential intrusions. 
It utilizes gathered threat knowledge and remains operational, even in the absence of updated information.
Based on its advice, the ITSs can dynamically and proactively adapt to recent threats to minimize and mitigate future intrusions from malicious adversaries.


Our goal is to reduce the risk linked to the exploitation of recently uncovered vulnerabilities that have not been classified and/or do not have a script to reproduce the exploit, considering the potential that they may have already been exploited as zero-day exploits.
Our experiments demonstrate that the proposed solution can effectively learn and replicate National Vulnerability Database's evaluation process with 99\% accuracy.
\end{abstract}


\section{Introduction}\label{sec:introduction}

The current threat landscape is escalating and is being marked by a proliferation of malicious cyber-attacks~\cite{a1,a2,a3,a4}, that encompass various forms, such as malware propagation, phishing, and ransomware, among others. 
Governmental and industrial entities are actively seeking more robust solutions to counter these evolving threats effectively.

Current solutions deploy a perimeter defense against intrusions and malicious adversaries. 
However, after a successful breach, most systems are left at the mercy of the intruder.
Established research has developed methods and techniques that go beyond perimeter defense, focusing on Intrusion Tolerant Systems (ITSs). 
These systems can tolerate intrusions by mitigating unauthorized access to critical information and/or avoid system exploitation after a successful breach.


ITSs follow the design pattern of replicated services.
These services can be replicated across multiple nodes or rotated periodically with a new node to refresh the exposed node. 
ITSs ensure both the \textit{liveness} and \textit{security} of a service by maintaining correct execution, even if performance is reduced, and by protecting the security of the data.
In addition, ITSs have been designed to dynamically adapt against multidisciplinary threats. 
This is particularly crucial given the ever-changing attack landscape where ITSs are deployed. 
A variety of techniques are used to adapt the system, including but not limited to, replica rejuvenation~\cite{bessani2008crutial, garcia2019lazarus}, diversity~\cite{bessani2008crutial, garcia2019lazarus, chun2008diverse, distler2011spare}, and N-version programming~\cite{distler2011spare}. 
These strategies collectively enhance the system's resilience and responsiveness to the dynamic threat landscape.

Yet, to further enhance adaptability, an ITS can also utilize information from public sources such as social networks, news, and online databases with information on vulnerabilities, exploits, and cyberthreats. 
This information allows the ITS to adjust proactively and bolster its defenses against identified 
cyberthreats, exploits, and vulnerabilities, by modifying configurations and operating systems (OSs) for increased resilience.
The concept of risk assessment and proactive adjustment was first proposed by the research of Heo et al.~\cite{heo2017designing}, and expanded by Garcia et al.~\cite{garcia2019lazarus}, both described and analyzed in Section~\ref{sec:related_work}.

These solutions collect information from online exploit and vulnerability databases, assess the risk to the current system using the base scores from the retrieved data, and propose a lower-risk alternative, if available. 
The trade-off in using these scores is that they require manual evaluation of each vulnerability before being assigned. 
Consequently, there is a significant delay between the discovery of new exploits or vulnerabilities, its risk assessment, and the evaluation of the ITS~\cite{ruohonen2019look}.

The process starts when a vulnerability is submitted for evaluation by the National Vulnerability Database (NVD). 
It is added to the Common Vulnerabilities and Exposures (CVE) database with only the description provided by the applicant~\cite{cveprocess}.
After adding to the CVE database, the new vulnerability starts the CVE Analysis process which will provide it with additional information for the CVE.


Due to the lengthiness of the process, because of the different participants, there is a time gap between reception and evaluation of a vulnerability, that is directly related with the threat of the vulnerability (i.e., the more severe the vulnerability, the slower the process) which can go up to a year~\cite{ruohonen2019look}.

Furthermore, at the time of writing, the backlog of vulnerabilities in NVD was significantly increasing because of the absence of support for evaluating new CVEs. 
Out of the 3,370 CVEs received in March 2024, only 199 underwent successful analysis~\cite{nvdbacklog}. 
If the challenge persists, the response time for addressing new vulnerabilities will escalate even further, potentially affecting the NVD CVE database dependability status as a source of information.



As such, the present work introduces HAL 9000 (HAL for short), a novel Machine Learning (ML) based Risk Manager.
It provides advice on secure and resilient configurations using publicly available Open-source Intelligence (OSINT) information, and  addresses the above-mentioned challenge, by automatically predicting the CVSS score of unrated CVEs, and reassessing the ITS configurations.

Considering the challenges faced by NVD, its primary innovation lies in employing ML techniques to assess new CVEs still in the NVD evaluation process. 
By predicting the CVSS score using the CVE description, HAL removes the dependency on the NVD score process, allowing it to take immediate action against new vulnerabilities.
State-of-the-art Risk Managers~\cite{heo2017designing, garcia2019lazarus} depend on the NVD CVE full evaluation, since they use the CVSS base score to assess the risk.

The purpose of HAL's CVSS score prediction tool is to provide a preliminary score for assessing a vulnerability while awaiting the completion of the NVD's evaluation process.
This temporary score results from the prediction step, which is the first step presented in Figure~\ref{fig:arch_tool}.
Once the NVD evaluation is concluded, HAL updates the scores on the predicted CVEs and reexecutes the configuration risk assessment function.

HAL's objective is to recommend secure configurations for ITSs, based on the knowledge gathered from OSINT online databases.
However, considering the environment where ITSs are deployed, and to safeguard HAL from potential malicious attacks, it is highly advisable for the ITS to provide a secure environment for its deployment. 
For instance, both Lazarus~\cite{garcia2019lazarus} and Skynet~\cite{freitas2023skynet} propose a two-plane architecture: a secured controller plane isolated from the environment and an execution plane exposed to potential threats. 


The objective of deploying a Risk Manager in a secured environment is to prevent malicious adversaries to compromise its correct operation.
This can occur either by interfering with the local OSINT database, in case of HAL, through input injection~\cite{malik2016database}, or attacking the ML model through data poisoning~\cite{yerlikaya2022data}.
As such, it is assumed that ITSs that employ Risk Managers are capable of providing a secure execution environment.

In summary, the main contribution of this paper is a novel Risk Manager, HAL 9000, that: 
\begin{itemize}
    \item improves the current state-of-the-art of Risk Managers
    \item fully automates the process of risk assessment and new configuration advisory, i.e., does not need human intervention after initialization
    \item facilitates interchangeability between ML/AI models (modularity)
    \item is capable of advising new configurations, even in the presence of unassessed exploits/vulnerabilities
\end{itemize}



The rest of the paper is organized as follows:
Section~\ref{sec:related_work} reviews the current state of the art in Risk Managers, CVSS score predictors, and text document clustering algorithms~\cite{singhal2017survey}.
Section~\ref{sec:design} provides a detailed description of HAL's architecture and workflow.
Section~\ref{sec:performance} presents the experimental setup and a comparative discussion of the results.
Finally, Section~\ref{sec:conclusion} discusses the main conclusions of this research and offers perspectives for future work.

\section{Related Work}\label{sec:related_work}
In this section, the current state of the art in Risk Managers is explored, with a focus on their roles in assessing and advising on the safest configurations. 
Additionally, advancements in CVE score predictors and machine learning techniques for text document clustering are reviewed~\cite{hotho2005brief}.

\subsection{Risk Managers} 

\noindent\textbf{Diversity Policy for Intrusion Tolerant Systems}~\cite{heo2017designing}, proposed a diversity policy to be employed by recovery-based ITSs. 
It retrieves the information on known software vulnerabilities and advises on combinations that minimize the risk of common vulnerabilities.
This approach generates all possible configurations and decides based on the minimum sum of the CVSS scores.

The architecture was designed for Web servers, where each server supports a service accessed via the Internet.
The proposed architecture comprises a cleansing group, a central controller and a processing group. 
The cleansing group houses the Virtual Machines (VMs) undergoing recovery and later used in the processing group. 
The processing group consists of VMs providing the designated Web service and are exposed to outside threats. 
The central controller coordinates the VMs (adding/removing from each group) and hosts the configuration selection algorithm, advising on the configuration. 

To protect the system from undetected attacks, the system applies a proactive recovery mechanism. 
A central controller periodically rotates the state of each VM, transitioning between active, cleansing, ready, and active states. 
During the transition from the cleansing to the ready state, the central controller executes a selection algorithm, where it applies the diversity policy to decide on a suitable secure configuration.

Yet, the drawback of employing the diversity policy method lies in its dependency on the NVD database score, prior to implementing any system alterations. 
This reliance can result in delays to adapt to the new vulnerabilities~\cite{ruohonen2019look}, because they were added to the CVE database and have not been evaluated by NVD, i.e., do not have a score attributed.

\noindent\textbf{Lazarus}~\cite{garcia2019lazarus} introduced an ITS architecture that adjusts system configurations to reduce vulnerability risks based on recommendations from its Risk Manager module. 
The research includes a novel CVE scoring method and uses ML algorithms to detect shared CVEs\footnote{By Lazarus definition, a CVE is shared between two or more OSs by the CVE vulnerability configuration field and/or by the description similarity with a CVE from another OS.} between replicas using the CVE description.

The new calculation method reevaluates the CVSS base score using the information present in the CVEs details.
This includes factors such as the age of the CVE, the existence of a patch, and if the vulnerability has already been exploited, as presented in the following equations\footnote{\textit{oldness\_threshold} is set to 365 days in the experiments; \textit{now} and
\textit{v.published\_date} return the current day and the day when the vulnerability was published, respectively.

\textit{v.patched} and \textit{v.exploited} are boolean parameters that indicate whether a vulnerability has a patch or has been exploited, respectively (0 for false, 1 for true).}.

\begin{equation}
\resizebox{\columnwidth}{!}{$
\textrm{score}(v) = \textrm{CVSS}(v) \times \textrm{oldness}(v) \times \textrm{exploited}(v) \times \textrm{patched}(v)
$}
\label{eq:1}
\end{equation}


\begin{equation}
\resizebox{\columnwidth}{!}{$
\begin{split}
\textrm{oldness}(v) = \max \Biggl( 1 - 0.25 \times \frac{(now - v.published\_date)}{oldness\_threshold}, 0.75 \Biggr)
\end{split}$}
\end{equation}

\begin{equation}
\textrm{patched}(v) = 0.5^{v.patched}
\end{equation}
 
\begin{equation}\label{eq:final}
\textrm{exploited}(v) = 1.25^{v.exploited}
\end{equation}

Another contribution by Lazarus was the detection, in the NVD database, of different CVEs with similar vulnerabilities affecting different applications/OSs.
This was discovered through description analysis. 
For instance, CVE-2014-0157 affects OpenSuse 13, CVE-2015-3988 affects Solaris 11.2, and CVE-2016-4428 affects Debian 8.0, but their description is similar, suggesting a common exploit across all OSs.

In order to analyze CVEs that follow the same pattern, Lazarus employed K-means, a clustering technique from ML, to verify their description.
K-means is an algorithm used for unsupervised learning that clusters the CVEs by their description similarity.

After clustering, in the evaluation phase, Lazarus's Risk Manager employs the new scoring system, penalizing pairs of replicas with common CVEs or same clustered CVEs.
Clustering suggests a likelihood of the exploit impacting both replicas.
Based on this evaluation, Lazarus recommends a more resilient system configuration against intrusions.

The Risk Manager used in Lazarus introduced a trade-off between security and resilience. 
It highlights that a system solely focused on minimizing risk may not be ideal for Byzantine Fault Tolerant (BFT) protocols. 
In a scenario where a pair of replicas can have one or more shared unpatched CVEs, it is possible to compromise the system by executing parallel attacks.
Even though the recommended configuration might not be the most secure, there is a guarantee that an unpatched vulnerability will not affect more than one replica at the same time, maintaining the BFT invariant of $3f + 1$.



Besides the drawback of being dependent on the NVD score system, Lazarus's Risk Manager also requires human intervention necessary to execute the K-means algorithm. 
The optimal number of generated clusters has to be manually visualized and inserted into the algorithm, which increases the execution time, and subsequently the vulnerability time window.
In addition, K-means is an algorithm susceptible to outliers~\cite{singh2013analysis}, i.e., the mean value of a cluster can be influenced by the outliers, which will affect the resulting clusters.
This can lead to wrong assessments.

\subsection{CVSS score prediction}\label{subsec:score_prediction}

With the increasing applications of Artificial Intelligence (AI) and latest advancements in Deep Learning, many processes can now be automated or replicated by computers. 
Tasks that are manually performed multiple times and follow a consistent pattern can be mimicked by AI and Deep Learning algorithms. 
This automation can speed up these processes, allowing operators to focus on refining and verifying the results, and providing feedback to improve the algorithms.

Given the lengthy nature of the evaluation process for new CVEs and the challenges faced by evaluators, researchers have explored new procedures to apply AI and Deep Learning algorithms that can automatically predict the CVSS scores for CVEs, enabling a quicker assessment.
As such, the related work on CVSS score predictors is addressed.




\noindent\textbf{Khazaei et al.~\cite{khazaei2016automatic}} introduced a method for predicting CVSS scores using natural language processing (NLP).
This method learns from previously available CVE vulnerabilities by analyzing their descriptions and associated CVSS scores.

During data preprocessing, the method applied text mining tools and techniques to extract feature vectors. 
The authors evaluated the application of three different algorithms for predicting CVSS scores: Support Vector Machines (SVM), Random Forest, and fuzzy systems.
Among these, the fuzzy systems provided the best accuracy, correctly scoring 88\% of the evaluated CVEs.

In conclusion, the study found that using automatic predictors can reduce human error and increase the speed of CVSS score calculation.


\noindent\textbf{Sahin et al.}\cite{sahin2019conceptual} extended the prior work of Han et al.\cite{han2017learning}, which predicted the severity levels of vulnerabilities based on their descriptions. 
In their study, they used similar feature extraction methods with word embeddings and prediction models using Convolutional Neural Networks (CNNs). 
Additionally, it incorporated Long Short-Term Memory (LSTM) networks and Extreme Gradient Boosting (XGBoost).

The objective of this research was to predict severity scores in addition to severity levels. 
The original work categorized vulnerability severity into ranges: low (0.1 - 3.9), medium (4.0 - 6.9), high (7.0 - 8.9), and critical (9.0 - 10.0).

For feature extraction, the authors removed words that only occurred once in the sentence corpus and trained the word2vec continuous skip-gram model, introduced by Mikolov et al.~\cite{mikolov2013efficient}. 
After generating feature vectors for each word, they converted the description sentences into vector representations by concatenating the word vectors.

The authors concluded that vulnerability descriptions contain valuable information that can be used with deep learning algorithms such as LSTM, CNN, and gradient boosting to predict severity scores with an average error of 16\%.


\noindent\textbf{Elbaz et al.~\cite{elbaz2020fighting}} proposed using CVSS vector prediction to address N-Day vulnerabilities.
The objective of CVSS vector prediction is to forecast the base metrics that constitute the CVSS standard and assess the severity of vulnerabilities. 
This approach uses linear regression on vulnerability descriptions to provide basic metrics for newly discovered vulnerabilities.

This method differs from previous work as it predicts the individual metrics rather than the overall score, allowing CVSS to offer more detailed information about the vulnerability and enhance the  ``explicability'' of the results.

For preprocessing, the authors used a bag-of-words approach on the vulnerability descriptions, followed by a filtering scheme to remove irrelevant words. 
Subsequently, a regression model was trained for each metric to be applied during assessment.

The authors concluded that while the method by Khazaei et al.~\cite{khazaei2016automatic} achieves higher accuracy for the base score, their approach offers better explainability.
They recommend implementing two pipelines: one for accuracy prediction and the other for ``explicability''.


\noindent\textbf{Costa et al.\cite{costa2022predicting}} proposed combining text preprocessing using NLP techniques with vocabulary expansion and the application of the Deep Learning method DistilBERT\cite{sanh2019distilbert}.
The objective of the research was to predict CVSS metrics using vulnerability descriptions, similar to the work of Elbaz et al.~\cite{elbaz2020fighting}.

For preprocessing, lemmatization and stemming were applied to the data.
Tokenization was performed using the Transformers library. 
The experiments tested the accuracy of combining the data with vocabularies of 5,000, 10,000, and 25,000 words added to the tokenizer's vocabulary.

The authors then tested several deep learning methods, specifically DeBERTa, BERT, ALBERT, and DistilBERT, to evaluate the accuracy and quality of the assessed data. 
The results showed that DistilBERT provided the most accurate predictions, while ALBERT was the least accurate.

The study concluded that DistilBERT is a state-of-the-art model for CVSS prediction, with enhanced performance when combined with lemmatization and a 5,000-word vocabulary.
However, no analysis was made on the accuracy for the base score since the main idea is to provide insight for the experts.


\noindent\textbf{Kai et al.\cite{kai2023vuldistilbert}} developed VultDistilBERT, a method to assess vulnerability severity, similar to the approach by Sahin et al.\cite{sahin2019conceptual}, using a distillation model. The method addresses data imbalance by augmenting data and using optimal subsets.

The data augmentation process generates more data from raw data without increasing its quantity. This is achieved by synonym replacement and random deletion to expand the sample set. The optimal subset selection involves choosing a subset of CVSS metrics during training to incorporate into the vulnerability descriptions, thereby enhancing the amount of textual information.

The method then uses the DistilBERT model as a text characterization tool, processing the preprocessed data. The resulting feature vectors from DistilBERT are then classified using a linear layer.

The authors concluded that their proposed method achieved state-of-the-art performance in vulnerability severity assessment, with a 97\% assessment accuracy.

\subsection{Text Document Clustering}\label{subsec:clustering}

Text clustering is a technique used in different areas of text mining and information retrieval.
Initially investigated to enhance the precision or recall in information retrieval systems~\cite{rijsbergen1979information, kowalski2007information}, as an efficient way of finding the nearest neighbors of a document~\cite{buckley1985optimization}, in browsing a collection of documents~\cite{cutting2017scatter}, or in organizing the results of a search engine response~\cite{zamir1997fast}.

Text Document Clustering is an unsupervised technique that groups documents into categories using unsupervised learning algorithms. 
According to Lazarus, online databases may contain entries with different identifiers that affect distinct systems but have identical descriptions, indicating the same vulnerability.
Applying text document clustering can mitigate this inaccuracy, enhancing the quality of data collected from online vulnerability databases. 
Therefore, the present subsection addresses the related work on Text Document Clustering.


\noindent\textbf{Steinbach et al.}~\cite{steinbach2000comparison} compared the results of applying agglomerative hierarchical clustering with K-means for document clustering. 
The study evaluated the implementations of the Unweighted Pair Group Method with Arithmetic Mean (UPGMA), K-means, and bisecting K-means.

To measure the quality of these algorithms, the authors used an internal quality measure, overall similarity, and an external quality measure, entropy. 
Overall similarity assesses cluster cohesiveness, while entropy measures the quality of the created clusters.

The authors concluded that ``given the linear run-time performance of bisecting K-means and the consistently good quality of the clusterings that it produces, bisecting K-means is an excellent algorithm for clustering a large number of documents''~\cite{steinbach2000comparison}.






\noindent\textbf{Singh et al.}~\cite{singh2011document} compared the performance of K-means, Heuristic K-means, and Fuzzy C-means for document clustering. 
The study explored different feature selection conditions (with and without stop words, with and without stemming) and various representations (term frequency, term frequency-inverse document frequency, and Boolean).

The trade-offs between the algorithms are as follows: K-means clustering quality is sensitive to initial seeds, Heuristic K-means adds computational cost due to the use of heuristics, and Fuzzy C-means does not produce hard clusters but provides a degree of membership for all created clusters.

For evaluation, the authors used internal and external criteria: residual sum of squares and purity, respectively. 
The results showed that Heuristic K-means performs better than K-means, but Fuzzy C-means is a more robust flat clustering algorithm.


\noindent\textbf{Mendonça et al.}\cite{mendoncca2019clustering} studied the effectiveness of classical literature clustering algorithms when applied to free text documents. 
They selected five clustering algorithms for their experiments, each capable of word-embedding document representation: K-means\cite{hartigan1975clustering}, Expectation–Maximization (EM) Clustering using Gaussian Mixture Models (GMM)\cite{liu2002document}, Spectral Clustering\cite{shi2000normalized}, Mean Shift~\cite{comaniciu2002mean}, and Density-Based Spatial Clustering of Applications with Noise (DBSCAN)~\cite{ester1996density}. 
The study aimed to observe the behavior of these algorithms in the task of document clustering.

For evaluation, the authors used Normalized Mutual Information and Homogeneity scores to measure the overall quality of the clustering solutions.

The results indicated that the chosen parameters greatly influence the performance of the algorithms. 
Among the selected algorithms, K-means performed the best.



\noindent\textbf{Asyaky et al.}\cite{asyaky2021improving} conducted an experiment to enhance the performance of density-based algorithms, specifically DBSCAN\cite{ester1996density} and HDBSCAN~\cite{mcinnes2017accelerated}.
The authors preprocessed the documents using lemmatization, stemming, and document embeddings to reduce dimensionality and improve the accuracy of these clustering algorithms.

For evaluation, they used the Adjusted Rand Index, which compares the pairs of objects in the resulting clustering to a ground truth clustering, and the Adjusted Mutual Information, which is based on information-theoretical mutual information~\cite{wagner2007comparing}.

The authors concluded that the obtained results surpassed most existing methods on the state of the art on the same subject.

\section{HAL 9000 Architecture}\label{sec:design}

This section provides an overview of HAL 9000's architecture and workflow.



\begin{figure*}[t]
  \centering
  \includegraphics[width=0.45\textwidth]{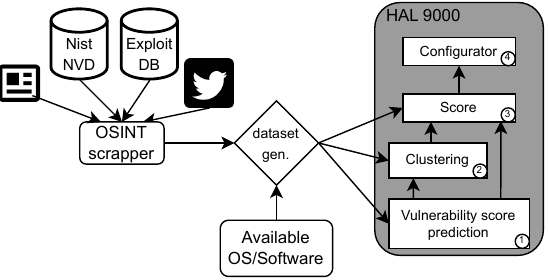}
  \caption{HAL 9000 Architecture and workflow, along with the integration of an OSINT scrapper tool. }
  \label{fig:arch_tool}
\end{figure*}

\noindent\textbf{HAL's architecture}: depicted in Figure~\ref{fig:arch_tool}, is constituted by four components: the Vulnerability score predictor, the Clustering algorithm, the Score reassessment, and the Configurator.

The Vulnerability score predictor is used in the assessment of new CVEs by providing a score using its description.
It uses an ML algorithm that is trained with the data of established vulnerabilities. 
Specifically, it uses the information present in the CVE's description and CVSS score metrics to train the model.
Afterwards, it assesses the new CVEs through their description and/or CVSS metrics depending on the applied algorithm, predicting the CVSS base score.
The score is then reassessed using the equations~\ref{eq:1} to \ref{eq:final}.

For CVSS score prediction, HAL was implemented and tested, with each contribution detailed in Subsection~\ref{subsec:score_prediction}. Based on the collected results presented in Section~\ref{subsec:results}, the method proposed by Khazaei et al.~\cite{khazaei2016automatic} demonstrated the best accuracy compared to the other algorithms.
However, during the architecture design phase, a modular approach was adopted to facilitate the interchangeability between different types of algorithms, allowing for future improvements.
This was achieved by restricting interaction between elements to the bare minimum, i.e., the dataset is transmitted to the element for training, thereafter accepting only the data to be evaluated.
Both datasets are passed in the form of a CSV file, containing the CVE id, description, and CVSS metrics.

The Clustering algorithm applied in the architecture follows the concept established by Lazarus.
It is possible for databases to have different CVE entries which affect different software/OSs but are similar in their vulnerability.
The objective of this component is to identify these occurrences through clustering using the CVE description.
Clustering is done through the use of an ML algorithm. 
This creates a set of clusters to classify each description based on their similarity.

For this component, the ML algorithm used should be fully automated to reduce the total execution time window and eliminate the possibility for human bias or error.


Furthermore, it is important for the algorithm to be robust to outlier data in order to prevent inaccurate clustering when inferring new data. 
Examples of ML algorithms that demonstrate insensitivity to outliers include, but are not limited to, HDBSCAN~\cite{mcinnes2017accelerated} and OPTICS~\cite{scikit-learn}.

Nonetheless, HAL's architecture was designed to accommodate other types of ML algorithms, i.e., supervised learning, semi-supervised learning, and deep learning. 
To allow this, the component follows the same modular design from the Vulnerability score predictor.

The Score reassessment component extends the contribution proposed from Lazarus.
It recalculates the CVSS score to more accurately reflect the CVE vulnerability.
As mentioned in section~\ref{sec:related_work}, Lazarus recalculates the score using the additional information in the CVE to improve its quality, by taking into account its age, patch availability, and occurrence of an exploit~\footnote{A vulnerability is considered exploited when there are reports of its occurrence or when code is provided to exploit the vulnerability.}.
HAL's architecture augments the Lazarus scoring system by considering the attention that the vulnerability receives in the ``wild''.

Lazarus equations make small adjustments to the CVSS base score, halving the score when a patch is available.
However, in practical terms, the existence of a patch does not guarantee its installation.
In 2022, incident responders were brought in to remediate attacks that began with exploited vulnerabilities, such as the ProxyShell and Log4Shell vulnerabilities. 
Both had existing patches at the time of compromise~\cite{unpatched}.

As such, HAL extends the calculation by readjusting the score to account for the potential non-installation of a patch and incorporating the weight of the Exploit Prediction Scoring System (EPSS)~\cite{jacobs2021exploit} score, as presented in equation~\ref{eq:hal}. 
The EPSS determines the exploit probability of a CVE in the wild.
Afterward, HAL calculates the \textit{hal\_score(v)}, as shown in equation~\ref{eq:hal}. 
This calculation involves determining the \textit{score(v)} with and without the patch application (if it exists), then applying the weight of the EPSS probability of being exploited and not exploited.



For example, the CVE-2017-11882 is an exploit with a score of 7.8 (considered High in NVD standards), that has a patch available. 
Under Lazarus calculations it will result in a new score of $3.65$, which is considered Low, by the Qualitative Severity Rating Scale~\cite{qualitative}. 
However, this does not reflect the current scenario of the exploit. 
This exploit was discovered in the year 2017, and even though it has a patch available, it is still an active threat, with an exploit probability of 97.99\%, based on the EPSS information. 
With the extension proposed by HAL, the new calculation scores at $7.2$, which is still High.
The new assessment considers a more sensible approach regarding the threat of a CVE, considering situations where the patch application might be delayed.




\begin{equation}
\resizebox{\columnwidth}{!}{$
\textrm{score}(v) = \textrm{CVSS}(v) \times \textrm{oldness}(v) \times \textrm{exploited}(v) \times \textrm{patched}(v)
$}
\label{eq:new}
\end{equation}


\begin{equation}
\resizebox{\columnwidth}{!}{$
\textrm{hal\_score}(v) = \textrm{score}(v) \times ( 1 - \textrm{EPSS}(v)) + \textrm{score}(v)_{wp} \times \textrm{EPSS}(v)
$}
\label{eq:hal}
\end{equation}


The final component, the Configurator, using the gathered information proposes a secure and resilient configuration for the ITS to deploy.
The component outputs two values, the \textit{security\_risk(config)} and the \textit{resilient\_risk(config)}.
Its objective is to minimize both outputs, prioritizing the \textit{resilient\_risk(config)}, i.e., shared CVEs calculation to avoid situations of ``break one, break all'', where ITS nodes are targeted by parallel attacks.
The \textit{security\_risk(config)} is calculated by summing the new assessed CVSS scores and predicted scores for new CVEs in a given configuration from all participating nodes.
The resilience calculation, the algorithm pairs every node in the configuration two by two and sums the CVSS score between shared CVEs and between CVEs clustered by their similarity.

\begin{equation}\label{eq:security}
\textrm{security\_risk}(config) = \sum_{n_i \in config} \sum_{v \in V(n_i)} \textrm{hal\_score}(v)
\end{equation}

\begin{equation}\label{eq:resilience}
\textrm{resil\_risk}(config) = \sum_{n_i, n_j \in config} \sum_{v \in V(n_i, n_j)} \textrm{hal\_score}(v)
\end{equation}

\subsection{HAL 9000 workflow}
HAL's workflow, depicted in Figure~\ref{fig:arch_tool}, illustrates the data path from OSINT data retrieval to the recommended configuration.

This subsection provides detailed information on HAL's workflow, describing the different paths taken by the data through its various components during execution.

HAL's workflow is divided into a four-step process, illustrated in Figure~\ref{fig:arch_tool}:
\begin{enumerate}
    \item Predicting CVE scores for unassessed CVEs.
    \item Clustering similar CVEs based on their descriptions.
    \item Performing risk assessment of the system configuration.
    \item Providing the most resilient and secure configuration.
\end{enumerate}

\noindent\textbf{1st step:} HAL receives a dataset of vulnerabilities and identifies CVEs that do not have assigned scores, i.e., CVEs that only have descriptions, with status ``Received''. 
For these unassessed vulnerabilities, HAL uses a Random Forest model to estimate scores based on evaluation patterns observed in the assessed vulnerabilities within the training dataset.


\noindent\textbf{2nd step:} The CVEs are clustered based on their descriptions to identify similar vulnerabilities.
The result of this process will later be used to calculate the resilience risk of the configuration, as presented in equation~\ref{eq:resilience}. 
This equation considers the shared vulnerabilities between nodes that are present in the CVE's configuration or through the description similarity from the clustering process.
This process avoids situations where different software are affected by distinct vulnerabilities, but in fact are the same vulnerability considering that both descriptions are similar.


\noindent\textbf{3rd step:} HAL then reassesses the scores of the vulnerabilities using previously established equations (refer to equations~\ref{eq:1} to~\ref{eq:hal}). 
This process is also applied to CVEs assessed by the Vulnerability Score predictor. 
New CVEs have the necessary information that enables the application of the new equations to the predicted score, specifically, the added date, if they were exploited, and the existence of a patch\footnote{Newly added vulnerabilities typically do not have patches for application.}.


\noindent\textbf{4th step:} After reassessing the scores of CVEs, HAL generates all conceivable variations of the given software/Operating System (OS) configurations. 
For each configuration, HAL calculates both its security level and resilience level using equations~\ref{eq:security} and~\ref{eq:resilience}, respectively. 
The resulting set of configurations is then arranged in order of resilience and security levels. The top configuration, representing the most resilient and secure option, is recommended to the ITS.

\section{Experiments}\label{sec:performance}
A total of three experiments were conducted to validate the proposed contributions.

The first experiment aimed to identify the best data preprocessing technique and clustering algorithm for a Risk Manager. 
Lazarus applied the bag of words (bow) technique for preprocessing and K-means for clustering, using the elbow method to find the optimal number of clusters (K). 
However, the research did not provide insights into the choice of algorithm, dataset size, selected K value, or how to handle situations where the elbow is difficult to identify, i.e., the point of inflection is less perceptive.
Considering the trade-offs of these techniques and the current state-of-the-art discussed in subsection~\ref{subsec:clustering}, two distinct preprocessing methods were considered for HAL: bag of words and embeddings. 
Subsequently, twelve different clustering algorithms were analyzed and applied to the dataset, each using one of the preprocessing techniques. 
After clustering, the results were applied to HAL's implementation to visualize the effects on its scoring system.

The second experiment validated HAL's new scoring system against established state-of-the-art methods. 
HAL's implementation was compared with Lazarus's Risk Manager and Heo et al.\cite{heo2017designing}, referred to as Heo for readability. 
The database initially contained CVEs until the end of 2022, simulating a deployment scenario at the beginning of 2023. 
To simulate the passage of time, a month's worth of CVEs was injected into the database during the experiment. 
Sixteen different operating systems were considered, as detailed in the Experimental setup subsection.

The third and final experiment evaluated the performance, Root Mean Square Deviation (RMSE), and accuracy of three different state-of-the-art methods for CVSS score prediction. 
Using our dataset, the accuracy and best approach to predict the score were analyzed, including CVSS metric prediction, CVSS severity prediction, and CVSS score prediction.



\subsection{Experimental setup}


For the experiments, a dataset containing information on CVEs, vulnerabilities, and exploits related to OSs and software was necessary. 
However, no existing tool or dataset provided the required information at the time of writing. 
To address this, a generator was developed to create a dataset with vulnerability information on various OSs and their installed software. 
To ensure a fair comparison with other Risk Managers, the dataset information was limited to entries from the NVD database and ExploitDB, as these were the sources used by the compared Risk Managers.

The generated dataset included 38,218 vulnerabilities from various versions of software produced by different manufacturers, encompassing Windows, Fedora, Ubuntu, CentOS, Debian, OpenSuse, Solaris, OpenBSD, and FreeBSD.
The dataset included information such as the CVE ID, publication date, last modification date, exploit description, exploit configuration, base metric version 2, and, if available, base metric version 3~\cite{mell2006common}.





The experiments were conducted on a virtual machine within the QEMU hypervisor, running Debian 12.
The virtual machine was equipped with 62 GB of RAM, a 32-core CPU, and an NVIDIA GeForce RTX 3090 graphics card with 24 GB of VRAM.


\subsection{Results and Discussion}\label{subsec:results}

This section presents the results of the experiments and provides a subsequent discussion of the findings. 
The results and respective discussion are organized in the order in which the experiments were conducted.

\begin{figure}[ht]
  \centering
  \includegraphics[width=\columnwidth]{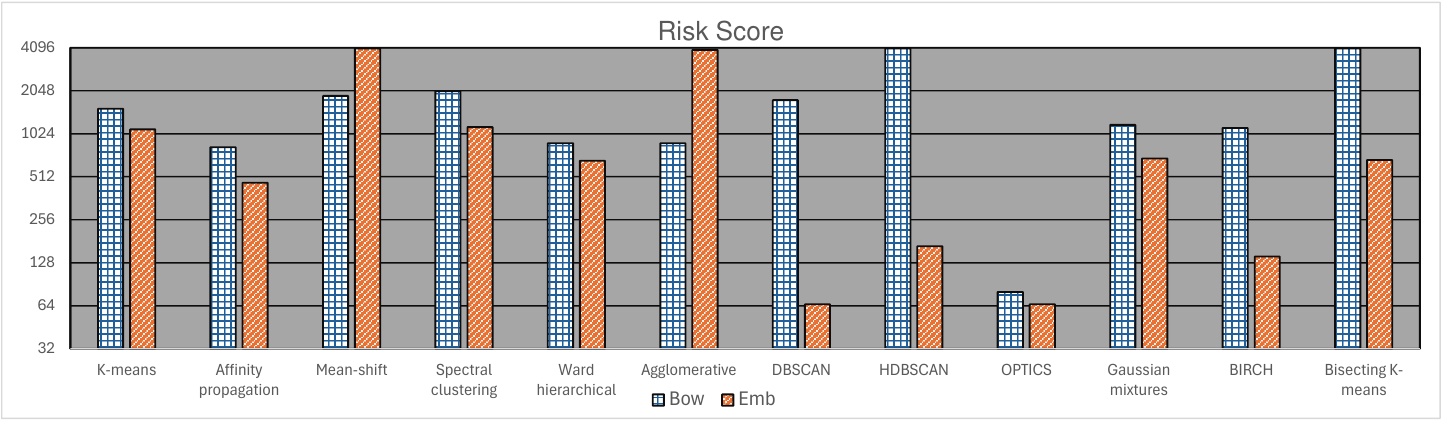}
  \caption{Evaluation of several clustering algorithms and subsequent effects on the HAL risk calculation (lower is better). In each algorithm two different approaches to preprocess data are applied, bag of words (bow) and sentence embeddings (emb).}
  \label{fig:eval}
\end{figure}

From the experiments conducted with various clustering algorithms, presented in Figure~\ref{fig:eval}, it was observed that sentence embeddings are the optimal preprocessing technique for the given data. 
This technique effectively maintains the relationships between words within sentences.

From Figure~\ref{fig:eval}, it is also possible to observe that OPTICS and DBSCAN yield the lowest risk scores compared to the other algorithms, based on the risk calculation from equation~\ref{eq:1}.
Both algorithms are insensitive to outliers and designed to be scalable, addressing the limitations of the K-means algorithm.

As such, taking into consideration these results, HAL utilizes the sentence embeddings for preprocessing data, and OPTICS for data clusterization.

\begin{figure}[ht]
  \centering
  \includegraphics[width=0.9\columnwidth]{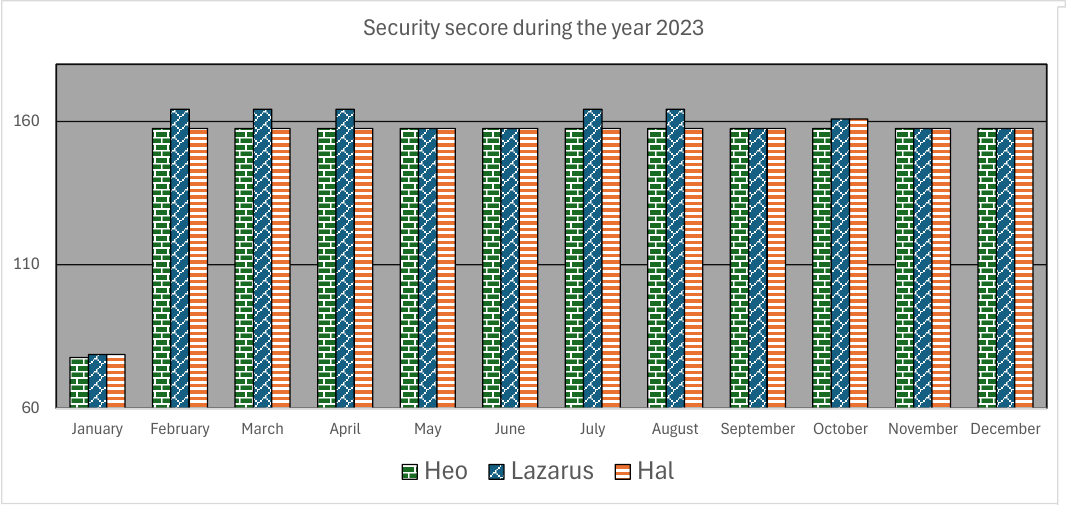}
  \caption{Evaluation of the different Risk Managers, considering the level of security, i.e., the sum of the CVSS score of each CVE present in the advised configuration (lower is better).}
  \label{fig:eval_security}
\end{figure}

\begin{figure}[ht]
  \centering
  \includegraphics[width=0.9\columnwidth]{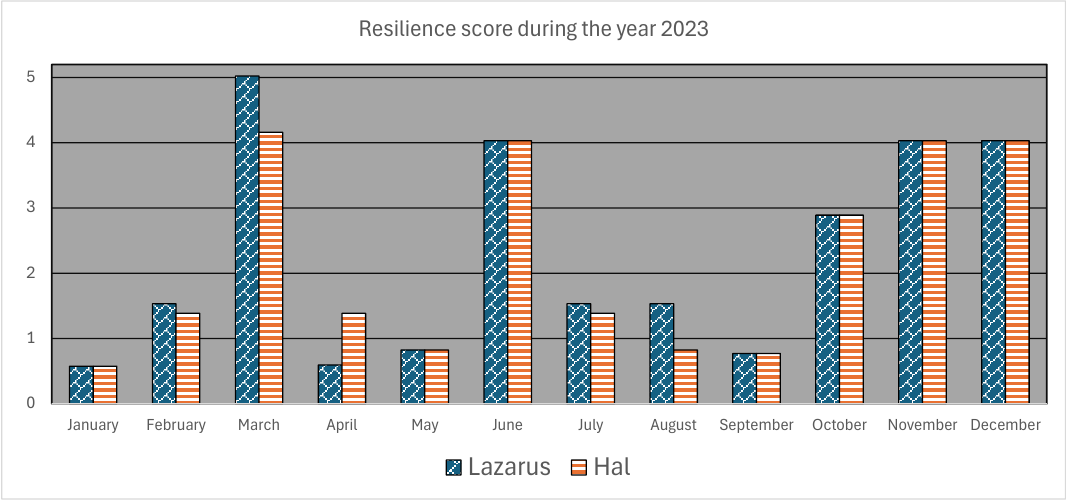}
  \caption{Evaluation of the different Risk Managers, considering the level of resilience, i.e., the multiplication between the reassessed CVSS score of the common CVEs and respective EPSS score (by shared CVE or by clustering) present in the advised configuration (lower is better).}
  \label{fig:eval_resilience}
\end{figure}

For the risk assessment, we provided two separate graphs for clarity, presented in Figures~\ref{fig:eval_security} and~\ref{fig:eval_resilience}.

The first graph shows the security score for each Risk Manager throughout 2023, while the second graph displays the resilience score\footnote{The resilience score reflects the number of shared vulnerabilities that can undermine an ITS.
Shared vulnerabilities between nodes increase the threat of parallel attacks.} for each Risk Manager over the same period.

Note that the resilience score for Heo is not calculated in the second graph because it was originally designed for non-replicated nodes. 
Although Heo's implementation was redesigned for the replicated scenario, it does not use a clustering algorithm.
As such, it is possible that the advised configuration and respective calculation might have hidden shared vulnerabilities which will not show up in its resilience score.

\begin{table*}[ht]
\caption{Analysis of different state-of-the-art AI implementations for CVSS prediction.}
\centering
\begin{tabular}{l c c c c c}
\hline
Model & Accuracy (\%) & RMSE & Execution Time (s) & Classifiers & Target Variable \\ \hline
Khazaei et al.~\cite{khazaei2016automatic} & 99 & 0.66 & 132.85 & Random Forest & CVSS score (interval) \\ \hline
Costa et al.~\cite{costa2022predicting} & 87 & 1.55 & 1432.14 & Linear & CVSS Metrics \\ \hline
VulDistilBERT~\cite{kai2023vuldistilbert} & 90 & 3.63 & 195.49 & Linear & CVE Severity \\ \hline
\end{tabular}
\label{tab:cvsspred}
\end{table*}

Both graphs use the Lazarus calculation method for the scores presented. 
During the experiments, each Risk Manager employed its technique to provide the best configuration. 
However, for comparison purposes, we applied the Lazarus scoring method since its a method that is closer to the other Risk Managers. 
Additionally, for the resilience comparison between HAL and Lazarus, we standardized the implementation by using the same preprocessing technique and clustering algorithm. 
The goal of this comparison was to demonstrate that HAL's score calculation system can provide a better configuration than Lazarus.

In the security graph (y-axis), the calculated values represent the sum of the CVSS scores recalculated using Lazarus equations. 
In the resilience graph (y-axis), the values represent the sum of the recalculated CVSS scores multiplied by the sum of the shared CVE EPSS, which indicates the probability of being exploited in the wild. 
This approach was used because a more active vulnerability (with a higher probability of exploit) can attack several nodes simultaneously, which is not considered in the security calculation.

Initial observations from both graphs indicate a trade-off: Heo provides better security, while Lazarus offers better resilience. However, HAL aims to combine both factors, providing configurations that are both resilient and secure. 
For example, in February, March, July, and August, HAL delivered configurations with lower risk that were both resilient and secure compared to Lazarus.

It is important to note that occasionally HAL produces configurations that are more secure but less resilient compared to Lazarus, as seen in April. 
This occurred because, in its settings, it was opted to choose more secured configurations over more resilient ones, considering that although the configuration might be exposed to parallel attacks, these would have a lower security score, i.e., have a lesser impact to the node compared with more severe vulnerabilities.

Table~\ref{tab:cvsspred} presents the experimental results for each of HAL's implementations with each state-of-the-art CVSS score predictors. 
Each implementation predicts a different target variable, as shown in the ``Target Variable'' column.

Although the predicted data is discrete in every implementation, it can be translated into continuous data to calculate the predicted score.
This allows for the computation of the RMSE of the predicted score, which is included in the table.

Khazaei et al.\cite{khazaei2016automatic} utilized traditional ML algorithms trained with the descriptions and scores of known CVEs to predict the score. 
Costa et al.\cite{costa2022predicting} applied deep learning algorithms trained with the descriptions and CVSS metrics of known CVEs to predict the CVSS metrics of new vulnerabilities. 
VulDistilBERT~\cite{kai2023vuldistilbert} employed deep learning algorithms trained with CVE descriptions and CVSS severity to predict the CVSS severity of new vulnerabilities.

The results, presented in Table~\ref{tab:cvsspred}, indicate that the method proposed by Khazaei et al. achieves the best accuracy with the provided dataset. 
Additionally, since it uses traditional ML methods (Random Forest), it requires less total execution time to train the model and infer new vulnerabilities. 
This implies a quicker response against new vulnerabilities in HAL's use-case scenario.

The method proposed by Costa et al., which predicts each CVSS metric separately, has an average accuracy of around 87\%.
Applying the CVSS v3.1 equations results in calculated CVSS scores with a higher RMSE compared to the original values. 
It is important to note that this method aimed to aid vulnerability reviewers in speeding up the assessment procedure for new vulnerabilities, which differs from HAL's objective.

Lastly, the method proposed in VulDistilBERT~\cite{kai2023vuldistilbert} achieves 90\% accuracy in predicting the severity of new CVEs. 
Although it has high accuracy for severity prediction, mapping it to actual CVSS scores is challenging since severity is relative and spans ranges: low (0.1 - 3.9), medium (4.0 - 6.9), high (7.0 - 8.9), and critical (9.0 - 10.0). 
This makes predicting precise scores difficult. 
For our calculations, we considered the highest value represented by each severity level to calculate the CVE base score, which explains the high RMSE.

After analyzing the results for each of HAL's implementations with a state-of-the-art CVSS score predictor, it is evident that the algorithm proposed by Khazaei et al.~\cite{khazaei2016automatic} achieves the highest accuracy compared to the other algorithms, resulting in a lower RMSE.


\section{Conclusion}\label{sec:conclusion}
The HAL 9000 is a Risk Manager capable of advising on secure and resilient configurations independently of the CVE rating system.
This independence is crucial in addressing highly dangerous vulnerabilities and quasi-zero-day exploits, where timely response is critical to preventing system compromise.

HAL's proactive awareness of new vulnerabilities and its advanced assessment capabilities give it a significant advantage over established risk managers. 
This enables ITSs to respond more quickly to new threats and rejuvenate exposed nodes, thereby enhancing overall system security and reliability.


For future work, we plan to enhance HAL's capabilities by incorporating data from penetration testing and automated testing tools. 
These tools will be integrated into a comprehensive framework alongside HAL. 
Additionally, we intend to deploy HAL in ITS implementations~\cite{freitas2023skynet, wang2003sitar} to evaluate its risk assessment performance in real-world scenarios and assess any potential system overhead.

We have also identified two important challenges that will be taken into account for our future work.
First, explore methods for scenarios where OSINT sources are either unavailable or compromised, which can impact negatively HAL's output. 
Second, develop a framework that balances the security and resilience levels, allowing administrators to make informed decisions.

\bibliographystyle{unsrt}
\bibliography{sn-bibliography}


\end{document}